\documentstyle[12pt,psfig,twoside]{article}
\begin{document}
\hbadness=10000
\pagenumbering{arabic}
\pagestyle{myheadings}
\markboth{J. Letessier, J. Rafelski and A. Tounsi}
{Evolution of the Quark--Gluon Plasma}
\title{Formation and Evolution of the\\ Quark--Gluon Plasma}
\author{$\ $\\
\bf Jean  Letessier$^1$, Johann Rafelski$^{1,2}$ {\rm and} Ahmed
Tounsi$^1$\\ $\ $\\
$^1$ Laboratoire de Physique Th\'eorique et Hautes Energies\thanks{\em
Unit\'e  associ\'ee au CNRS UA 280.}\\
Universit\'e Paris 7, 2 place Jussieu, F--75251 Cedex 05.
\\ \\
$^2$ Department of Physics, University of Arizona, Tucson, AZ 85721\\
and Theory Division, CERN, 1211 Geneva 23, Switzerland\\
}
\date{}   % Deleting this command produces today's date.
\maketitle
%%%%%%%%%%%%%%%%%%%%%%%%%%%%%%%%%%%%%%%%
\begin{abstract}
{\noindent
Imposing an equilibrium between the thermal pressure of deconfined quarks
and gluons and the dynamical compression pressure exercised by in-flowing 
nuclear matter, we study the initial thermal conditions reached  in a 
quark--gluon plasma fireball formed in a relativistic heavy ion collision. 
We show that entropy is produced primarily in the pre-equilibrium stage 
of the reaction. 
We test our approach, comparing our results with the S$\to$W/Pb  
collision results at 200 GeV A  and 
find a surprising degree of agreement assuming about 50\% stopping. 
We apply our method to a determination of the conditions in collisions 
of Au$\to$Au at 11 GeV A and  Pb$\to$Pb at 157 GeV A, assuming full stopping 
of momentum, energy and baryon number. 
Our detailed results directly determine the spectral shape 
and abundance of (strange) hadrons and  electromagnetic probes (photons, 
dileptons) produced in the collision, and we explore specific 
experimental consequences. 
}\end{abstract}
\centerline{Published in Phys. Lett. B333 (1994) 484}
\vfill
\noindent {CERN-TH.7304/94} \\{PAR/LPTHE/94--23}\\  
\noindent {May 1994}\newpage 
%%%%%%%%%%%%%%%%%%%%%%%%%%%%%%%%%%%%%%%%%%%%%%%%%%%
\noindent{\bf 1. Introduction}\\

Our objective is to obtain a simple and qualitative understanding of the
initial conditions reachable in high-energy collisions of nuclei. Although the
information about the high density state that we observe 
using hadronic probes is originating in the late stages of the evolution
of the high-density matter formed in relativistic nuclear collisions, some 
basic physical principles will allow us to go back in time in order to
obtain a rather precise picture of the initial state. 
These considerations allow us to determine
the expected formation rates of electromagnetic probes of the dense state
(dileptons, photons), which provide an important verification of our
understanding of the structure of high-density matter.  In our present 
investigation, which is based on the 
hypothesis that the quark--gluon deconfined phase has been formed 
in the interaction, we shall use a slight generalization to chemical
(particle abundance) non-equilibrium of the quark--gluon plasma (QGP) 
equations of state described by us previously. These incorporate perturbative 
QCD interaction effects and thermal particle masses \cite{Let94b}. We have 
already shown \cite{Let93,Let94,Let94a} that the formation of the QGP phase 
is indeed, at the highest now available energy of 200 GeV A,  the 
simplest hypothesis capable to consistently account for all experimental 
data. At the lower end (10--15 GeV A) of the energy range considered here,  
this QGP hypothesis is not inevitable, given the present-day data sample 
\cite{Let94c}. 
The key step we make in the present work is to establish, based on very 
plausible and simple dynamical conditions, the early values of temperature 
and chemical potentials, which in our previous work arose from the study 
of experimental results.

Implicit in the physical picture employed here 
(seeRefs.\,\cite{Let94,Let94a} for more details), is that 
a space-time region of hot, dense hadronic matter with nearly thermal 
properties is formed in the centre of momentum (CM) frame (`central region'), 
which can be characterized by the statistical parameters, the temperature $T$ 
and the baryo-chemical potential $\mu_{\rm B}=3\mu_{\rm q}$ or equivalently 
the quark fugacity $\lambda_{\rm q}=\exp(\mu_{\rm q}/T)$, and similarly 
for strange quarks $\lambda_{\rm s}=\exp(\mu_{\rm s}/T)$.   In addition, 
we will study the approach to chemical equilibrium (particle-abundance 
equilibrium) by all different components.  We thus introduce the chemical 
occupancy factors of the different particles:  
gluons $ \gamma_{\rm G}\,$, light quarks and antiquarks $\gamma_{\rm q}\,$, 
and strange quarks $ \gamma_{\rm s}$. 
When the accessible phase space is saturated we have $\gamma_i=1$ and the 
chemical equilibrium is established. Note that 
the $q\,\ ,\bar q$ abundance is controlled by the two parameters 
$\lambda_{\rm q}$ and $\gamma_{\rm q}$ and thus the number of both can be 
established independently: the fugacity factor 
$\lambda_{\rm Q}=\gamma_{\rm q}\lambda_{\rm q}$ 
determines the quark abundance, and the factor  
$\lambda_{\overline{\rm Q}}=\gamma_{\rm q}\lambda_{\rm q}^{-1}$ 
determines the antiquark abundance. We  assume that  
thermalization  is  faster than the chemical equilibration --- 
but also note that the mechanism causing the rapid formation of a 
thermal particle distribution remains little understood. 
Studying final particle 
spectra it can be convincingly argued in favour of thermalization. An 
extensive analysis \cite{ULI93} has been carried out accounting
for diverse distortions of the spectra caused by the disintegration of 
unstable hadrons and longitudinal flow.  These effects in the  
transverse mass $m_\bot$ particle spectra  are small
at high $m_\bot$; in the domain $m_\bot>1.5$ GeV  
one can directly infer for S--W/Pb collisions at
200 GeV A a source temperature of about $232\pm5$ MeV in the spectra of 
several strange-particles \cite{WA85new} as well as in the spectra of 
$\pi^0$ and $\eta$ \cite{WA80spec}.

Let us briefly summarize the four arguments pointing to the QGP nature of the 
dense fireball formed in collisions at 200 GeV A:\\
\phantom{\bf ii}{\bf i)} {\bf Strange-quark fugacity:}
$\lambda_{\rm s}\sim1$ is natural for a directly disintegrating QGP phase where
the symmetry between the $s$ and $\bar s$ quarks is reflected naturally in the
value of $\lambda_{\rm s}=1$\,. In the confined phase, whatever the equation of 
state, $\lambda_{\rm s}=1$ is an exceptional condition at finite baryon density,
since strange baryons (carriers of $s$-quarks) and mesons  
(carriers of $\bar s$-quarks) have different masses.
Studying the relative abundance of strange-particles 
it is possible to precisely determine the final-state value of 
$\lambda_{\rm s}$. 
The analysis of the S$\to$W \cite{Let94,Raf91} and  S$\to$S 
\cite{Sol93} data has shown that the strange-quark fugacity 
$\lambda_{\rm s}\simeq 1$. The recently reported 
$\overline{\Omega}/\Omega$ result \cite{WA85OM} has provided 
further independent evidence \cite{Let94} that $\lambda_{\rm s}\simeq 1$. 
Note that one  finds, for the lower energy AGS results 
obtained with 14.6 GeV~A projectiles  \cite{Let94c,RD94}, a value 
$\lambda_{\rm s}=1.72\pm 0.19$, 
distinctly different from unity. \\
\phantom{\bf i}{\bf ii)} {\bf Strange-quark phase-space occupancy:}
Comparing the abundance of particles with unequal numbers 
of strange-quarks, one can obtain a measure of the strange-quark phase-space  
occupancy, and at 200 GeV A one finds $\gamma_{\rm s}\to 1$, suggesting that 
rather effective  processes (presumably gluon-based \cite{Raf93}) were 
available to produce strangeness \cite{Let94,Raf91,Sol93}. \\
{\bf iii)} {\bf Excess particle multiplicity/entropy:}
The flow of charged hadrons can be used \cite{Let93} to establish the hadronic 
multiplicity produced per participating baryon, and this quantity can 
be related to the entropy produced in the interaction. This entropy content 
is easier to bring into consistency with the QGP picture than with a more 
conservative reaction picture in which colour is frozen, thus, when other
conditions are equal, less entropy is produced. \\
{\bf iv)} {\bf Consistency with QGP equations of state:}
Using the QGP equations of state one finds \cite{Let94b} that the 
temperature in  $m_\bot$ spectra and $\lambda_{\rm q}$, as determined using 
strange-particle multiplicity, are consistent with the energy content 
available in  the collision. 

In this work we will be able to go beyond this last observation and to 
determine the observable values of $T,\lambda_{\rm q}$ alone on the basis 
of the CM collision energy and its stopping.  We will thus determine:\\
$\bullet$ the primordial temperature for the S$\to$W/Pb collisions 
at 200 GeV A ($E_{\rm CM}\simeq8.8$ GeV),\\and the conditions:\\
$\bullet$  prevailing in collisions Si$\to$Au at 14.6 GeV A, Au$\to$Au 
at 11 GeV A ($E_{\rm CM}\simeq2.6$ GeV),\\
$\bullet$  expected for the forthcoming Pb$\to$Pb collisions at
157 GeV A ($E_{\rm CM}\simeq8.6$ GeV).\\
In principle all our results are now fully ab initio and do not contain any
parameters, apart from the assumtion that, for S$\to$W/Pb, 
the stopping $\eta_i$ of energy, momentum and  baryon number is about 
50\% \cite{stop}, while we take
$\eta_i\simeq 1$ in the other cases considered above.

From the relaxation-time constants applicable to the strange-quark production
\cite{Raf93}, which is, without doubt, slower than for gluons 
and light quarks, 
one can infer that the time $t=t_{\rm s}=3$--$5$ fm/$c$.
At this time (nearly) complete chemical equilibrium is 
reached and the temperature has cooled to its observable value  
$T\vert_{t_{\rm s}}\equiv T_0$. 
Here $t_{\rm s}$ is about a factor 6--10 longer than what is found
for the relaxation time of the chemical equilibration of light quarks and 
gluons \cite{Gei93,Shu93}.  Correspondingly, we shall assume that in the 
epoch ($t=t_{\rm ch}\simeq 1$ fm/$c$) of the hot matter evolution,  when the  
quark and gluon  chemical equilibrium characterized by phase-space occupancy 
$\gamma_{\rm q}\vert_{t_{\rm ch}}\simeq\gamma_{\rm G}\vert_{t_{\rm ch}}
\simeq 1$ is reached, the  occupancy of the strange quark phase-space is about 
$\gamma_{\rm s}\vert_{t_{\rm ch}}=0.15$\,. 
We also presume that the initial thermalization time $t=t_{\rm th}$,
for quarks and gluons already made, is faster than the
chemical equilibration. We will take the chemical occupancy factor 
$\gamma_{\rm q}\vert_{t_{\rm th}}\simeq \gamma_{\rm G}\vert_{t_{\rm th}}
\simeq 0.2\,,\gamma_{\rm s}\vert_{t_{\rm th}}=0.03$ 
when considering the very initial  conditions of thecentral fireball. 
It is not possible to really start with all $\gamma_i=0$
since even in the pre-thermal phase a number of quark pairs and gluons will be
produced. On the other hand, the exact values of $\gamma_i$ do not matter since
conservation laws and other constraints almost fully determine the system.
Thus any other `reasonable' initial choice would in no way alter the results we
present here.

The energy (and baryon number) required in the fireball is drawn from the
fraction $\eta_{\rm E}$ of in-flowing energy and the fraction $\eta_{\rm B}$ of
in-flowing baryon number, which is contributing to the thermal central fireball.
These `stopping parameters' characterize the phenomenon of compression and
conversion of energy from relative motion into the thermal energy content of
the central fireball, and ultimately into the final particle multiplicity. It
is important to notice that the duration of the collision is considerable on
the scale applied here. A large nucleus has a diameter of $\simeq 12$ fm and
at the CM energy per nucleon available today,  up to 9.6~GeV~A, the 
Lorenz contraction factor ($E_{\rm cm}/m\simeq10$) implies that the collision 
time $t_{\rm col}\ge 1$ fm/$c$. The parton cascades \cite{Gei93} 
(albeit for higher energies) and QCD gluon multiplication
rates \cite{Shu93} place the time at which the chemical equilibrium of light 
quarks and gluons is reached at about $t_{\rm ch}=0.5$--$1$ fm/$c$ in agreement 
with what one could infer from simple perturbative considerations of QCD cross 
sections. 
Thus for {\it present-day} heavy-ion collisions ($E_{\rm beam}\le 200$~GeV~A) 
the thermalization and chemical equilibrium of quarks and gluons is reached
while the nuclear collision is taking place.  
During this collision period, the pressure exercised on the central fireball 
by the impacting matter is nearly constant. 
Therefore the degree of compression of the central fireball can be found by 
equating the dynamical pressure associated with the kinetic motion of the 
colliding nuclei, with the dominant resisting force, which is the internal 
{\it thermal} pressure of the fireball. Moreover, since the chemical 
equilibration of quarks and gluons occurs during the collision, it also occurs 
in condition of constant pressure.

Once the collision is terminated the central fireball approached the
chemical equilibrium of light quarks and gluons at some 
constant pressure will commence to cool down by producing the strange-quark 
pairs and also by expanding in all directions. Let us first imagine that 
there is no strangeness production. Then the expansion of a nearly 
{\it ideal} quantum gas would occur without production of additional entropy, 
considering a hydrodynamic expansion of a locally equilibrated system. 
Consequently, the specific entropy per baryon $S/B$ will remain nearly constant,
even in the presence of considerable particle evaporation: such emission 
processes are likely to reduce thebaryon and entropy content of the
fireball at comparable rate. Since the specific entropy is dimensionless, it
can only depend on dimensionless quantities, and for chemically
equilibrated system  with negligible strangeness fraction this can only be
$\lambda_{\rm q}=\exp(\mu_{\rm q}/T)$. Consequently, this expansion of 
the fireball in the vacuum will occur at fixed $\lambda_{\rm q}$\,. 
This situation is in our opinion little changed if one
considers the production of strange pairs; the entropy of the non-strange
fraction will remain nearly constant for the reasons described, and thus again
$\lambda_{\rm q}= $ const. The total entropy increases by the amount produced
making strange-quarks, while the energy transfer to the strange-quark gas 
reduces the speed of the collective expansion \cite{Cool}.

We will next discuss the dynamical pressure in the collision, and review,
in the following, in more quantitative manner the time scenario
described here, including a discussion of the pertinent statistical fireball
properties as well as the evolution of the entropy in the nuclear reaction. 
Some experimental consequences will also be studied.\\

%%%%%%%%%%%%%%%%%%%%%%%%%%%%%
\noindent{\bf 2. Pressure balance}\\

We now consider quantitatively the balance between the dynamical compression
exercised by the in-flowing matter and the resisting thermal pressure of the
constituents  of the central fireball. The pressure due to kinetic motion 
follows from well-established principles, and can be directly inferred, e.g. 
from Eq.\,(15) of  Ref.\,\cite{deG80} for the energy-momentum tensor:
\begin{equation} \label{Tdegrand}
T^{ij}(x)=\int\! p^iu^j f(x,p){\rm d}^3\!p\,,\quad i,j=1,2,3\,,
\end{equation}
where  $u^j=p^j/E$. We take for the phase-space distribution of incident 
particles $f(x,p)=\rho_{0}(x)\delta^3(\vec p-\vec p_{\rm CM})$. 
To obtain  the pressure exercised by the flow of matter, we consider 
the pressure component $T^{jj}$, with $j$ being  
the direction of $\vec v_{\rm CM}$. The result is:
\begin{equation}\label{Pdyn}
P_{\rm dyn}=\eta_{\rm p} \frac{p_{\rm CM}^2}{E_{\rm CM}}\rho_{0}\,.
\end{equation}
Here it is understood that the energy $E_{\rm CM}$ and momentum $P_{\rm CM}$
is given in the CM frame, and per nucleon.  
We have introduced  in Eq.\,(\ref{Pdyn}) the momentum stopping fraction
$\eta_{\rm p}$ --- only this fraction  $0\le\eta_{\rm p}\le 1$ of the 
incident CM  momentum can be used  by a particle incident on the central 
fireball  (the balance remains in longitudinal, unstopped motion) in order
to exercise dynamical pressure. For a target 
transparent to the incoming flow, 
there would obviously be no pressure exercised. 
In principle, Eq.\,(\ref{Pdyn})
defines the momentum stopping $\eta_{\rm p}$ 
and it is our hope and expectation
that this $\eta_{\rm p}\simeq \eta_{\rm E}$, the latter being defined by 
studying the fraction of the original beam energy found in the transverse 
direction to the beam motion \cite{stop}. 
The magnitude of momentum (energy) stopping has been
studied for different reactions.  In the S$\to$W/Pb 
reactions at 200 GeV~A, the 
energy stopping is $\eta_{\rm E}\simeq0.5$, and we therefore assume a similar 
value for our parameter $\eta_{\rm p}$.  At lower energies we will move towards 
full stopping, that is $\eta_i=1$. 
We also anticipate that $\eta_i=1$ when we 
study collisions of largest nuclei such as Pb$\to$Pb.
 
Note that the magnitude of the dynamical pressure can be obtained 
qualitatively as follows: 
assume that only a fraction  $0\le\eta_{\rm p}\le 1$ of the incident CM 
momentum can be used  by a particle incident on the central fireball 
in order to exercise pressure. 
We divide this change in momentum per increment in time 
and surface to obtain the pressure, converting the arriving particles 
into impact density and, taking also  $dt=dz/v_{\rm CM}$, we find again 
Eq.\,(\ref{Pdyn}). It is interesting to note that (with full stopping 
and using the normal nuclear density, $\rho_{0}=0.16\,{\rm fm}^{-3}$) 
one finds the benchmark value $P_{\rm dyn}=0.5$ GeV/fm$^3$ (respectively 
$ 1$  GeV/fm$^3$) at $\gamma_{\rm CM}=3.4$ (respectively $ 6.4$). 
In a symmetric collision this corresponds to an
incident projectile at 23.6 GeV/c$^2$ A (respectively $= 86.3$ GeV/c$^2$ A).

The idea of deconfinement and  QGP is to some extent based on the 
simple picture of a confined hadron
in which  the internal Fermi pressure of constituents is balanced
by the external vacuum pressure, which has the magnitude ${\cal B}\simeq
0.1\,{\rm GeV/fm}^3$. Applying the same idea to the determination of the
initial conditions in the collision, we demand that the internal  thermal  
(and Fermi) pressure $P_{\rm th}$ be balanced by 
the dynamical and vacuum pressure. We thus obtain the condition:
\begin{equation}\label{Pbal}
P_{\rm th}(T,\ldots)=P_{\rm dyn}+P_{\rm vac}=
	\eta_{\rm p}\frac{E^2_{\rm CM}-m^2}{E_{\rm CM}} \rho_{0}+{\cal B}\,.
\end{equation}

The QGP equations of state will determine the left-hand side of 
Eq.\,(\ref{Pbal}). These equation follow from a slight generalization of our 
previous work \cite{Let94b}, as we now allow for the absence of  
chemical equilibrium of gluons and light quarks. 
In short, we take ideal quantum gases and shall allow for the
suppression of the occupancy of gluons $G$ and all quarks by the factors 
$\gamma_{i}, i={q,s,G}$. We will allow for `thermal' masses of all 
these particles with $m_i^2=m^2_0+(cT)^2$\,. In  principle we have 
$c^2\propto\alpha_{\rm s}$, but given the
current uncertainty regarding the value of the coefficient $c$ we
shall simply explore its consequence in the domain  $c\sim 2$, arising
for $\alpha_s\sim 1$ in the standard formulas. We take
$m^0_{\rm q}=5$ MeV, $m_{\rm s}^0=160$ MeV, and $m_{\rm
G}^0=0$. We correct the quantum degeneracy factors using 
the perturbative  thermal QCD  corrections according to the
prescription of Ref.\,\cite{Let94b}.  
For the QCD coupling, we take $\alpha_s=0.6$. 
Numerical integration of the Bose/Fermi distributions for
quarks/gluons including these  effects allows us to obtain any
physical property of the QGP.

In Fig.\,\ref{Fini1} we show, in the $T$--$\mu_{\rm B}/T$ plane and at 
the chemical equilibration time $t_{\rm ch}$ of light quarks and 
gluons, i.e. for $\gamma_{\rm q}=\gamma_{\rm G}=1, \gamma_{\rm s}=0.15$, 
the QGP trajectories at constant energy per baryon (solid curves), 
beginning with $E/B=2.6$ (right curve), rising in increments of 1 GeV 
up to $E/B=10.6$ GeV (left curve).
Along these lines the QGP pressure varies. Assuming
that the stopping of energy and baryon number is practically the same 
$\eta_{\rm E}\simeq\eta_{\rm B}$ we have $E_{\rm CM}=E/B$. 
One can now find points of pressure balance implied by Eq.\,(\ref{Pbal}) 
for each CM energy, which are shown by the dashed lines in   Fig.\,\ref{Fini1}.
The highest assumes $\eta\equiv\eta_{\rm p}=\eta_{\rm E}=\eta_{\rm B}=1$, the 
lowest is for $\eta=0$ (vacuum pressure only), in increments of 0.25. The
intersections of these lines correspond to the temperatures $T_{\rm ch}$ 
attained at the end of the nuclear collision and 
before the transverse expansion 
and most of strangeness production occur.\\
%%%%%%%%%%%%%%%%%%%%%%%%%%%%%%%%%%

\noindent{\bf 3. Temperature evolution in the collision}\\

We now consider in more detail the different stages of the nuclear collision.
The particles observed in the collision are emanating at some low, freeze-out
temperature $T_{\rm f}$. Even though the spectra of high $m_\bot$ strange and
non-strange-particles are showing the same slope \cite{WA85new,WA80spec}, it
cannot be assumed that they have frozen out at the same time, and hence it is
convenient to introduce the strange-particle freeze-out temperature $T_{\rm
s}$. This can be the case, since  $T_{\rm f}\le T_{\rm s}$ could be indeed much
smaller than the temperature $T_0$ visible in the transverse particle spectra:
the collective flow velocity $v_{\rm f},v_{\rm s}$ superposes a Doppler blue
shift with:
\begin{eqnarray} 
T_{\rm f}=\sqrt{1-v_{\rm f}\over 1+v_{\rm f}} T_{0}\,,\qquad
T_{\rm s}=\sqrt{1-v_{\rm s}\over 1+v_{\rm s}} T_{0}\,.\label{doppler}
\end{eqnarray}
One finds $T_0\simeq232$ MeV for both strange and non-strange-particles at high
$m_\bot$, where other effects that distort the spectra (e.g. resonance
disintegration) become negligible. This observed temperature $T_0$ is
representative of the conditions of the fireball prior to its expansion after
the initial compression, but after strangeness flavour was produced.

We have already introduced two other temperatures: 
$T_{\rm ch} $ is the temperature
reached in the system after the chemical equilibration of light quarks and
gluons, but prior to the significant saturation of the strangeness phase-space,
the production of strangeness being slower than of light quarks. The 
temperature $T_{\rm th}$ is the temperature at which we can hypothesize that
thermal motion is established already, while the valence quarks, gluons and
quark--antiquark pairs produced are still far from chemical equilibrium. We
thus have the following temperature hierarchy. Here the time increases 
from left to right while temperature decreases:
\begin{equation}
T_{\rm th} > T_{\rm ch} >  T_0 \ge T_{\rm s} \ge  T_{\rm f}
\end{equation}
The first inequality is related to cooling during the approach to chemical
equilibrium (number equilibrium) of $q,\bar q, G$, the second one is due to the
cooling while  strangeness is produced and reaches its chemical
equilibrium, and the last two arise from  transverse
expansion in which thermal energy is transferred to the collective flow.

We are thus equipped with a hierarchy of times and temperatures and a fully
constrained time evolution scenario --- 
for a given set of three chemical equilibration 
parameters $\gamma_{\rm G},\gamma_{\rm q},\gamma_{\rm s}$ 
there are in principle two unknown statistical
parameters, $T$ and $\lambda_q$, which are determined after the
equations of state and with two additional constraints: in our case, either 
given $E/B$ and total pressure $P=P_{\rm dyn}+P_{\rm vac}$, or fixed $E/B$ and 
$\lambda_{\rm q}$. We have already given the resulting values of $T_{\rm ch}$
in Fig.\,\ref{Fini1}.  In Fig.\,\ref{Fini2} we show the effect of cooling 
due to the formation of strange-quark pairs. We show the quantity 
$\Delta T=T_{\rm ch} - T_0$ as a function of $E/B$, solid line is for full
stopping, the dashed one for $\eta_i=0.5$. These results were
obtained by computing at a given $E/B$ and 
the suitably associated fixed $\lambda_{\rm q}$, as implied by the results
of Fig.\,\ref{Fini1}, and increasing  the value of $\gamma_{\rm s}$ from
0.15 to 1. Similarly, we can compute from the values $T_{\rm ch}$ the
rise to $T_{\rm th}$ by exploring the effect of reducing  
$\gamma_{\rm G},\gamma_{\rm q}$ at constant pressure, but since the value of
$T_{\rm th}$ results form a chosen initial set of values for the pre-equilibrium
occupancy factors, we will not discuss in detail the resulting first 
thermal temperature, except to 
say that it is up to 150 MeV greater than $T_{\rm ch}$. 

\begin{table}[tb]
\caption{Conditions in different collision systems at 
different evolution stages.}
\begin{center} 
\begin{tabular}{|c||c|c|c|c|} 
\hline
Phase-&&\multicolumn{3}{|c|}{\phantom{$\displaystyle\frac{E}{B}$}$E/B$ [GeV]
\phantom{$\displaystyle\frac{E}{B}$}}\\
\raisebox{1mm}{space}&\raisebox{2mm}{$<\!s-\bar s\!>=0$} &2.6 & 8.8 & 8.6 \\
\cline{3-5}
occupancy&$\lambda_{\rm s}\equiv 1$ &$\eta=$ 1 &$\!\eta\!=\!
0.5\!$&$\eta=$ 1 \\
&&Au--Au&S--Pb&Pb--Pb\\
\hline\hline
&$T_{\rm th}$ [GeV]&0.260&0.410&0.471\\
$\gamma_{\rm q}=$ 0.2&$\lambda_{\rm q}$&9.95&1.78&2.00\\
&$n_{\rm G}/B$&0.20&1.55&1.25\\
&$n_{\rm q}/B$&3.00&5.12&3.77\\
$\gamma_{\rm G}=$ 0.2&$n_{\bar{\rm  q}}/B$&0.00&2.12&0.77\\
&$n_{\bar{\rm s}}/B$&0.02&0.16&0.13\\
&$\!P_{\rm th}\!$ [GeV/fm$^3$]&0.46&0.79&1.46\\
$\gamma_{\rm s}=$ 0.03&$\rho/\rho_0$&3.34&1.70&3.18\\
&$S/B$&11.8&40.0&33.4\\
\hline
&$T_{\rm ch}$ [GeV]&0.212&0.280&0.324\\
$\gamma_{\rm q}=$ 1&$\lambda_{\rm q}$&4.16&1.49&1.61\\
&$n_{\rm G}/B$&0.56&2.50&2.08\\
&$n_{\rm q}/B$&3.11&5.16&4.62\\
$\gamma_{\rm g}=$ 1&$n_{\bar{\rm q}}/B$&0.11&2.16&1.62\\
&$n_{\bar{\rm s}}/B$&0.05&0.25&0.21\\
&$\!P_{\rm ch}\!$ [GeV/fm$^3$]&0.46&0.79&1.46\\
$\gamma_{\rm s}=$ 0.15&$\rho/\rho_0$&3.35&1.80&3.19\\
&$S/B$&12.3&41.8&34.9\\
\hline\hline
&&$\gamma_{\rm s}$ = 1&$\gamma_{\rm s}$ = 0.8&$\gamma_{\rm s}$ = 1\\
\hline
&$T_0$ [GeV]&0.190&0.233&0.270\\
$\gamma_{\rm q}=$ 1&$\lambda_{\rm q}$&4.16&1.49&1.61\\
&$n_{\rm G}/B$&0.56&2.50&2.09\\
$\gamma_{\rm G}=$ 1&$n_{\rm q}/B$&3.11&5.12&4.60\\
&$n_{\bar q}/B$&0.11&2.12&1.60\\
$\gamma_{\rm s}=$ 0.8&$n_{\bar{\rm s}}/B$&0.28&1.27&1.07\\
or&$\!P_0\!$ [GeV/fm$^3$]&0.33&0.47&0.84\\
$\gamma_{\rm s}=$ 1&$\rho/\rho_0$&2.41&1.05&1.81\\
&$S/B$&14.1&49.5&41.7\\
\hline
\end{tabular} 
\end{center} 
\end{table}

This hierarchy 
of thermal conditions in the fireball can be seen for the cases of 
particular interest in table 1. 
In the first column we show the assumed values of the chemical
occupancies corresponding to the different stages. In the second, we list the
properties we shall present explicitly (when possible these are 
given per baryon, with
$\rho/\rho_{0}$ being the baryon density in units of the equilibrium
nuclear density). In 3rd, 4th and 5th columns we give our results 
for the three cases of
specific interest: first for the Si$\to$Au and Au$\to$Au 
collisions at $E_{\rm CM}=2.6$~GeV, 
$\eta_i=1$ (corresponding to the condition at 
BNL--AGS with 11--15 GeV A projectiles), then for the case of 
S$\to$W/Pb at $E_{\rm CM}=8.8$ GeV, $\eta_i=0.5$ (corresponding
to  S projectile penetrating a tube of matter in a central impact on a
heavy target at 200~GeV~A) and finally Pb$\to$Pb collisions at 
$E_{\rm CM}=8.6$ GeV, $\eta_i=1$
(corresponding to  Pb projectiles at 157 GeV~A colliding
at small impact parameter with Pb target).

In the first, top part of table 1, we present the
most hypothetical results, as it corresponds to
the earliest `th' period, just when we can apply statistical methods and
speak of temperature. Thus the value of the temperature $T_{\rm th}$ 
shown there is a direct consequence of the number of gluons and light 
antiquarks assumed here to be present (note that we always make sure that 
for each stopped baryon there are three quarks per baryon in the fireball). 
Next, in the second part of the table comes the epoch,
beginning with the temperature $T_{\rm ch}$, 
when the fireball reaches chemical equilibrium for light quarks 
and gluons, the penetration of nuclei ends, and the expansion and 
equilibration of strangeness abundance commences.  In the last, bottom
part, we consider the properties after strangeness has been produced,
for two slightly different values of $\gamma_{\rm s}$. 
As suggested by the S$\to$W results \cite{Let94}, we allow  a 
a slight deviation from the equilibrium in this case.  
We recall that in the transverse spectra the temperature $T_0$ is
visible at high $m_\bot$ and we see that the statistical  properties 
determined here are in remarkable agreement with the  values extracted 
from experiments:
$T_0=232\pm 5$ MeV, $\lambda_{\rm q}=1.48\pm 0.05$ \cite{Let94}.\\

%%%%%%%%%%%%%%%%%%%%%%%%%%%%%%%%%%%%%%%%%%%%%%%%%%%
\noindent{\bf 4. Some experimental consequences}\\

When looking at the Au--Au predictions at the lowest energy considered here 
(assuming QGP formation), the first and
most striking  result is that, at some early stage (first part of
table 1), there are practically only valence quarks present practically no
noticeable pair or glue production has taken place. Even
when we consider the chemically equilibrated system  at later time, we
see that the glue abundance is just equal to the number of $s$ and $\bar s$
quarks. This relatively small number of gluons may be inhibiting the
production of strangeness, and thus the assumed strangeness chemical
equilibrium, even if QGP were formed in these collisions, is 
probably over-optimistic. Because of the relatively large expected 
value of $\lambda_{\rm q}$, a simple test for QGP in this system is 
the measurement of the ratios of particles such as 
$\overline{\Lambda}/\bar p \simeq \overline{\Xi^-}/\overline{\Lambda}
\propto\gamma_{\rm s}\lambda_{\rm q}$, which should thus considerably exceed 
expectations \cite{Let94c}.
Also note that our earlier analysis of the Si--Au results \cite{Let94c}
suggested that, should QGP be formed in the 14 GeV A interactions, its 
disintegration would require a considerable re-equilibration 
in the final state,
since $\lambda_{\rm s}\simeq 1.7$ was found in the final state, and not 
$\lambda_{\rm s}\simeq 1$ as expected in rapid QGP dissociation. 
The necessary presence of the intermediate re-equilibration era complicates 
the arguments for, or against, possible  formation of QGP in these collisions.
We further note that the production of exotic strange matter 
containing an unusual strangeness fraction is difficult, and indeed unlikely, 
since $s/q\simeq 0.09$. We also note that even though this
ratio rises to 0.25 at the higher energies, 
there is so much more entropy produced there that it is hard to see how the 
system could cool down without 
dissociating by evaporating the excess entropy. 
Our results thus do not encourage searches for strange matter exotica
in relativistic heavy-ion collisions.

We next consider, in  table 1, the subtle differences between the
S$\to$Pb and Pb$\to$Pb collisions. We  predict a rise in 
temperature and  $\lambda_{\rm q}$ --- the observable temperature rises from 
$T_0=233$~MeV to $T_0=270$ MeV, which should indeed be easily visible 
in the high $m_\bot$-spectra.  Because of the considerable 
increase in baryon density (from 1.8$\rho_0$ to 3.2$\rho_0$)
we see a noticeable drop in specific entropy. This is intuitively
correct, since the energy per baryon is similar; 
the greater energy content per final-state particle at higher 
temperature thus entails a smaller number 
of particles per baryon, and hence a smaller entropy content per baryon. 
We also note that even the slight 
rise in $\lambda_{\rm q}$ has a considerable 
impact on  the ratio $\overline{\Lambda}/\Lambda$\,, which
is reduced by the factor $(1.49/1.61)^4=0.73$ 
due primarily to a greater number of
$\Lambda$, given the greater baryon density achieved. The specific yield of
strange-particles is also expected to drop, 
but it remains relatively high as the number of $\bar s$ is greater than the 
number of $\bar u$ or $\bar d$ (note
that $\bar q=\bar u+\bar d$, and naturally $\bar s=s$). Thus the ratios 
of particles such as 
$\overline{\Xi^-}/\overline{\Lambda}$, which are considered as
a test for the QGP phase \cite{Raf82,Raf93}, will increase by
35--50\% as we move to the Pb$\to$Pb system from S$\to$Pb. Interestingly,
this ratio will modestly increase while collision energy decreases, as 
long as the QGP phase is formed, due to an increase in the value 
of $\lambda_{\rm q}$, which we predict. A sudden decrease should follow, when
$E_{\rm CM}$ drops below the threshold for the formation of QGP.

We now comment on the consequences of the relatively 
high values of $T_{\rm ch}=280$ MeV (S$\to$Pb case) and 
$T_{\rm ch}=324$  MeV (Pb$\to$Pb case) and even greater 
pre-chemical-equilibrium values reported in table 1. 
A calculation of the dileptons yields in similar 
conditions was already performed \cite{Kat92}, and we have interpolated 
these results to the value of temperature 
applicable for an S$\to$Pb collision. 
In Fig.\,\ref{Fini3} we show (solid line) as a function of the dimuon
invariant mass the sum of the thermal 
QGP dimuons (short-dashed contribution), 
the hadron contribution (long-dashed component) and
the Drell--Yan (with $K=2$) together with renormalized $J\!/\!\psi$ 
contributions (dotted line, chosen to fit the $J\!/\!\psi$ peak). 
The relative yield of the QGP radiance, in the particularly 
interesting region between 1.4 and 2.6 GeV of the dimuon
invariant mass, arises from these normalizations and is hence not arbitrary. 
This result is in agreement with the experimental ones 
results presented recently \cite{Maz93} and shown here as open squares.  
While the agreement in slope between the 1.4 and 2.6 GeV 
dimuon invariant mass is due to the  magnitude of the source temperature,
the agreement in normalization must be seen as being probably a happy 
coincidence: 
the strength of the source is roughly proportional to the product of the 
abundance of quarks and antiquarks, which is dominated by the valence quarks, 
see table  1. However, the theoretical dilepton work \cite{Kat92} 
did not allow for any stopping of baryon number. 

The rise in temperature, seen as we move from the S$\to$Pb to the Pb$\to$Pb 
system, suggests a greater visibility of photons and dileptons in the heavier 
system. However, the situation is more complex: the rate of signal to noise 
of photons and dileptons is proportional to $n_{\rm q}n_{\bar{\rm q}}/S$, which 
drops by 20\% as we move from S$\to$W/Pb to Pb$\to$Pb collisions. Thus while 
the invariant-mass region of current interest 
could appear flatter due to higher temperature, 
the relative strength of the signal to hadronic
background  rate will be a bit less. More vexing is that this very 
interesting signal is much less specific and sensitive to the source 
conditions --- 
we hope to return in the near future to a more detailed study of the dilepton 
and photon spectra using the evolution, as determined here, of the 
QGP phase.\\ 

%%%%%%%%%%%%%%%%%%%%%%%%%%%%%%%%%%%%%%%%%%%%
\noindent{\bf 5. Final remarks}\\

Another  important result to observe in table 1 is that the entropy content, 
which determines the final-particle multiplicity \cite{Let93,Let94},
evolves very little and is to all purposes already present at the initial
stage, when  quarks and gluons are still far from abundance equilibrium 
\cite{Cool}. Practically all the rise we present is due to the formation 
of the strange flavour against the background 
of the thermalized and expanding  
quark--gluon fireball. We have an enormous amount of strangeness, as the 
tables show, and hence the fact that 20\% 
of entropy is due to strangeness production is not 
surprising. In total for the S$\to$Pb and Pb$\to$Pb, there are 
$n/B=0.25\, S/B$ particles per baryon, once chemical equilibrium is reached, 
in agreement with the `rule' that in a relativistic non-degenerate system 
on average, each particle carries $\simeq$ 4 units of entropy. 
The Au$\to$Au system is essentially a degenerate 
quark matter and one finds therefore less entropy, since
only quarks at the Fermi surface contribute. The entropy per
particle drops, and hence the coefficient given above 
is greater (0.25$\to$0.32).

Our present results reconfirm the key message regarding 
entropy \cite{Cool},  which is 
that it is produced  during the thermalization phase, 
and not during the approach to chemical 
equilibrium when most particles are produced. Thus the 
final hadronic multiplicity is
determined in the initial instants of the collision, 
and is not determined by model calculations, which  assume
initial condition.
 
As the system  evolves towards the final freeze-out conditions, its entropy 
content remains in essence unchanged, as we have so far 
discovered no entropy-generating mechanisms acting at this very 
last stage of the evolution.
Thus, irrespective of the values of freeze-out temperatures $T_{\rm s}$ 
and $T_{\rm f}$ at which the strange and, respectively,  non-strange 
particles decouple, the specific entropy content of the 
hadronic system is expected to remain the same.
Interpreting particle abundances in terms of 
chemical potentials and temperature of a source, and 
assuming an equation of state, the entropy 
content can be calculated. For the S$\to$W collisions 
at 200~GeV~A we found earlier \cite{Let93,Let94}
 that for the QGP equations of state  we should expect 50 units 
of entropy per baryon in the final state, as we now
also find in an ab initio calculation, as shown in table 1.
If the (multi-)strange (anti-)bar\-yons \cite{WA85new,WA85OM}, 
which served us in the determination of the properties of the fireball,
were to originate from a confined hadronic state, with properties described 
by a mixture of hadronic resonances (Hagedorn gas), we find an
entropy content that is nearly half as large as that of a QGP source. 
At a late time in collisions all hadrons have materialized. 
Unless some new physics is introduced that generates entropy in 
the evolution of the hadronic matter, in particular after 
strange-particles were produced from such a hadron gas, the situation is that 
the expected final particle multiplicity  must be very different for the 
two evolution scenarios, i.e. HG or QGP. Studying the hadronic 
multiplicity, we have determined that it implies an entropy-rich 
source \cite{Let93,Let94}. 

In the present work we have shown that this entropy-rich state is governed 
by the equations of state of the perturbative QGP. In particular 
we found (see table 1) in an ab initio calculation
that when (nearly) full chemical equilibrium is reached, the apparent 
source temperature and baryo-chemical conditions are as
deduced from  the analysis of the strange-particle data  for 200 GeV~A 
collisions. We have also demonstrated that the observed 
excess of dileptons in the invariant mass range 
1.4--2.6 GeV is consistently described by our QGP  
fireball model. 
We have explored the properties of a QGP fireball formed possibly 
in the Au$\to$Au collisions at 11 GeV~A and we have also
made detailed predictions about the conditions
expected in the forthcoming Pb$\to$Pb collisions at 200 GeV~A.\\

\vspace{0.5cm}
\noindent{\bf Acknowledgement}\\ 

J. R. acknowledges partial support from the US DOE grant DE-FG02-92ER40733.

A. T. wishes to thank Siki Ibn Mouloud for his poetic considerations on
{\sl Strangeness}. 
\vspace{0.3cm}
\vfill\eject
%%%%%%%%%%%%%%%%%%%%%%%%%%%%%%%%%

%\vfill\eject
%%%%%%%%%%%%%%%%%%%%%%%%%%%%%%%%%%%
\begin{figure}[h]
\vspace*{2cm}
\centerline{\bf FIGURE CAPTIONS}
\caption[]% Fig 1
{
The $T$--$\mu_{\rm B}/T$ plane. 
solid lines: trajectories at constant energy per baryon in QGP, with
$E/B=2.6$  up to $ 10.6$ GeV (rising in intervals
of 1 GeV from right to left); 
dashed lines: lines of constant pressure P in QGP with the value of P 
determined by Eq.\,(\ref{Pbal}) for any given $E/B$, with 
$\eta_{\rm p}=0$ up to $ 1$ (rising in intervals of 0.25 with $\eta$).
\label{Fini1}
}
\caption[] %Fig 2
{
$\Delta T=T_{\rm ch} - T_0$ as function of $E/B$, obtained at 
fixed $\lambda_{\rm q}$ with $\gamma_{\rm s}$ changing from 
0.15 to 1\,. Solid line $\eta_{\rm p}=1$, dashed line  $\eta_{\rm p}=0.5$\,.
\label{Fini2}
}
\caption[] %Fig 3
{
Spectrum of dimuons as a function of the dimuon invariant mass (arbitrary 
normalization), after Ref.\,\cite{Kat92}. The solid line is the sum 
of the thermal QGP dimuons (short-dashed contribution),  the 
hadron contribution (long-dashed component) and the Drell-Yan 
(with $K=2$) together with normalized $J/\!\psi$ 
contributions (dotted line, chosen to fit the $J\!/\!\psi$
peak). Experimental results (open squares) are read from Fig. 5, 
in \cite{Maz93}. 
\label{Fini3}
}
\vfill
\end{figure}
%\end{document} %TO PRINT FIGURES PERCENT THIS COMMAND

\newpage
\begin{figure}[p]
\vspace*{-4cm}
\centerline{\hspace{0.2cm}\psfig{figure=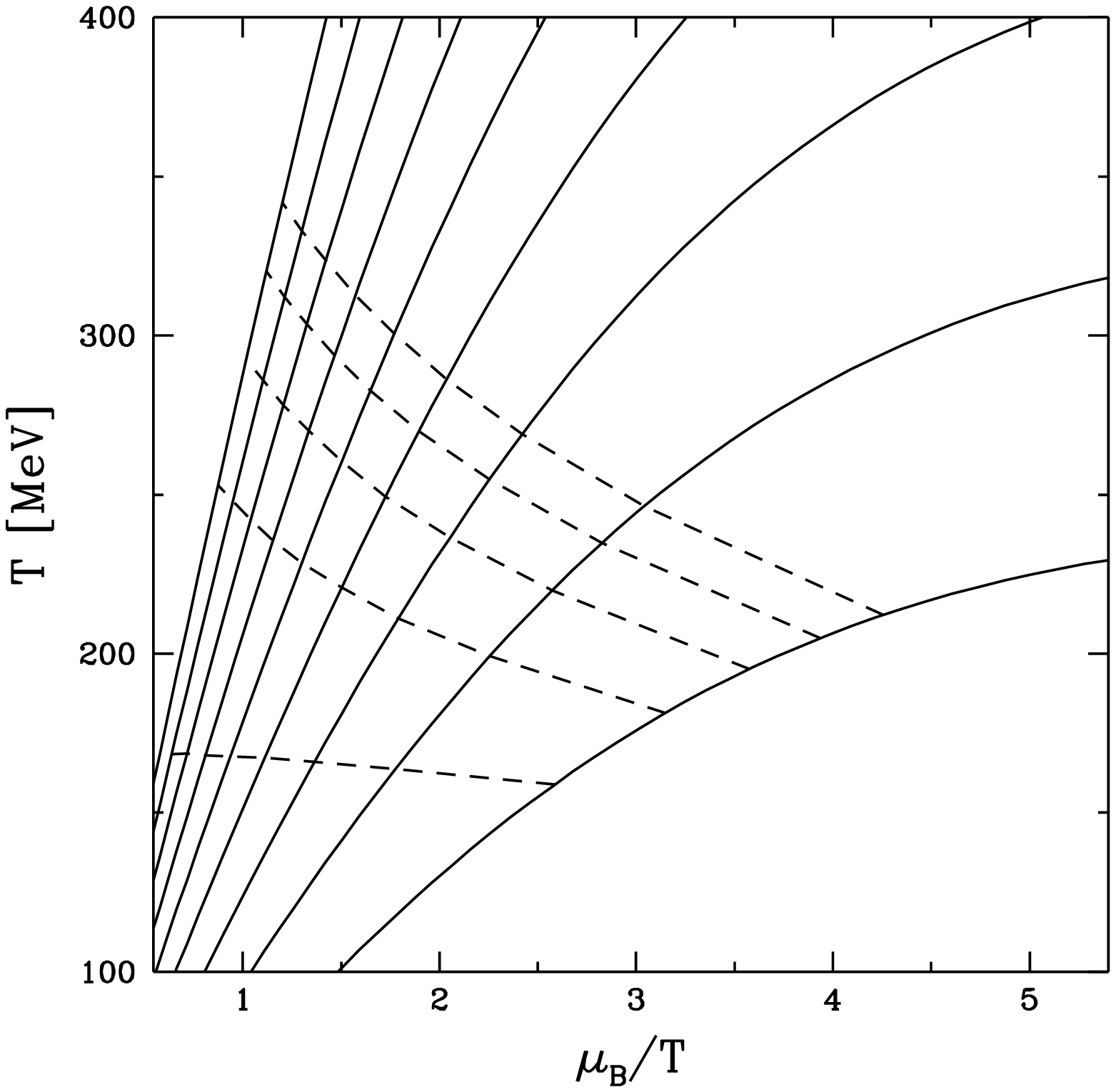}}
\vspace*{-2cm}
\centerline{\bf FIGURE 1}
\end{figure}
\newpage
\begin{figure}[p]
\vspace*{-8cm}
\centerline{\hspace{0.2cm}\psfig{figure=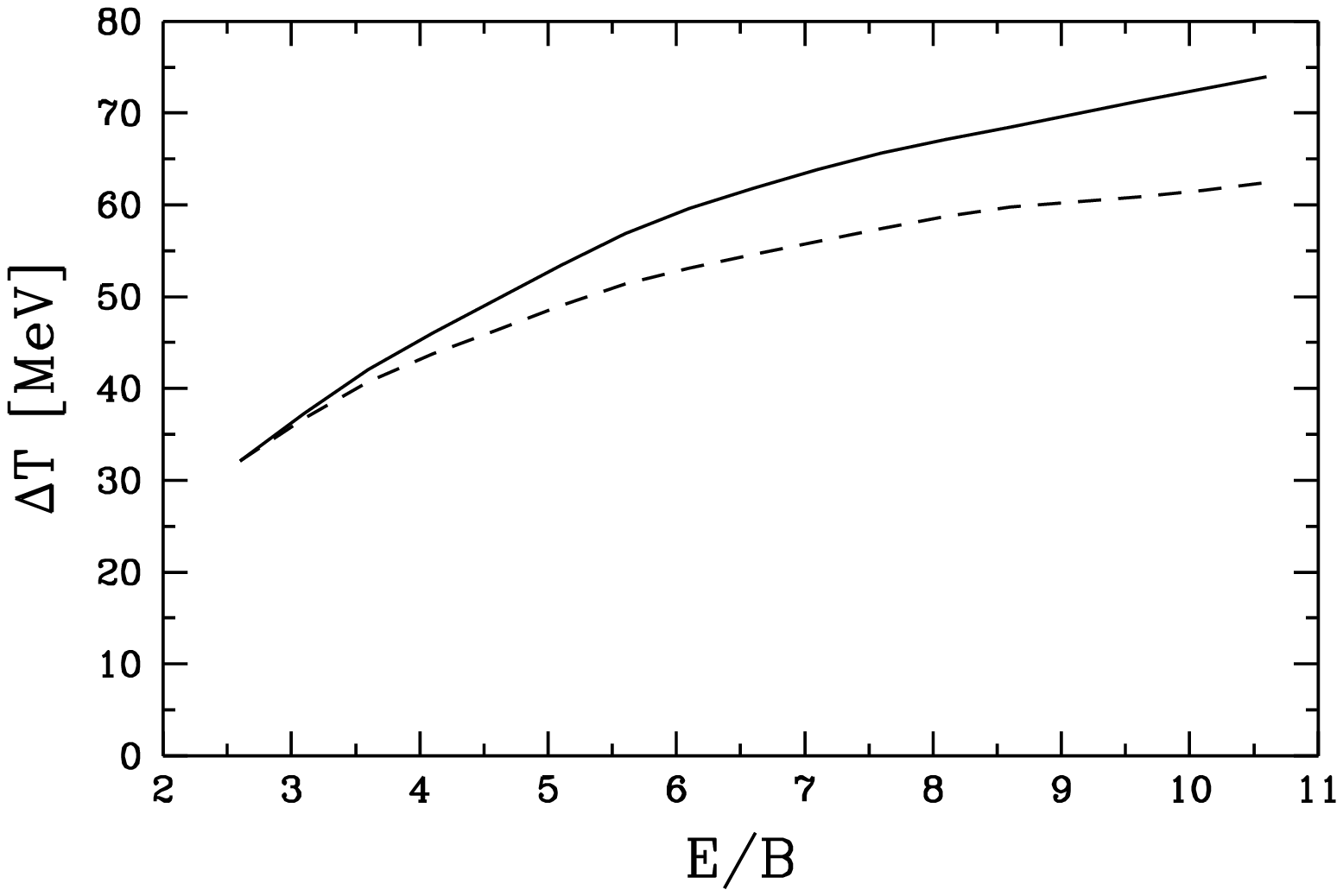}}
\vspace*{-2cm}
\centerline{\bf FIGURE 2}
\end{figure}
\newpage
\begin{figure}[p]
\vspace*{3cm}
\centerline{\hspace*{8cm}\psfig{figure=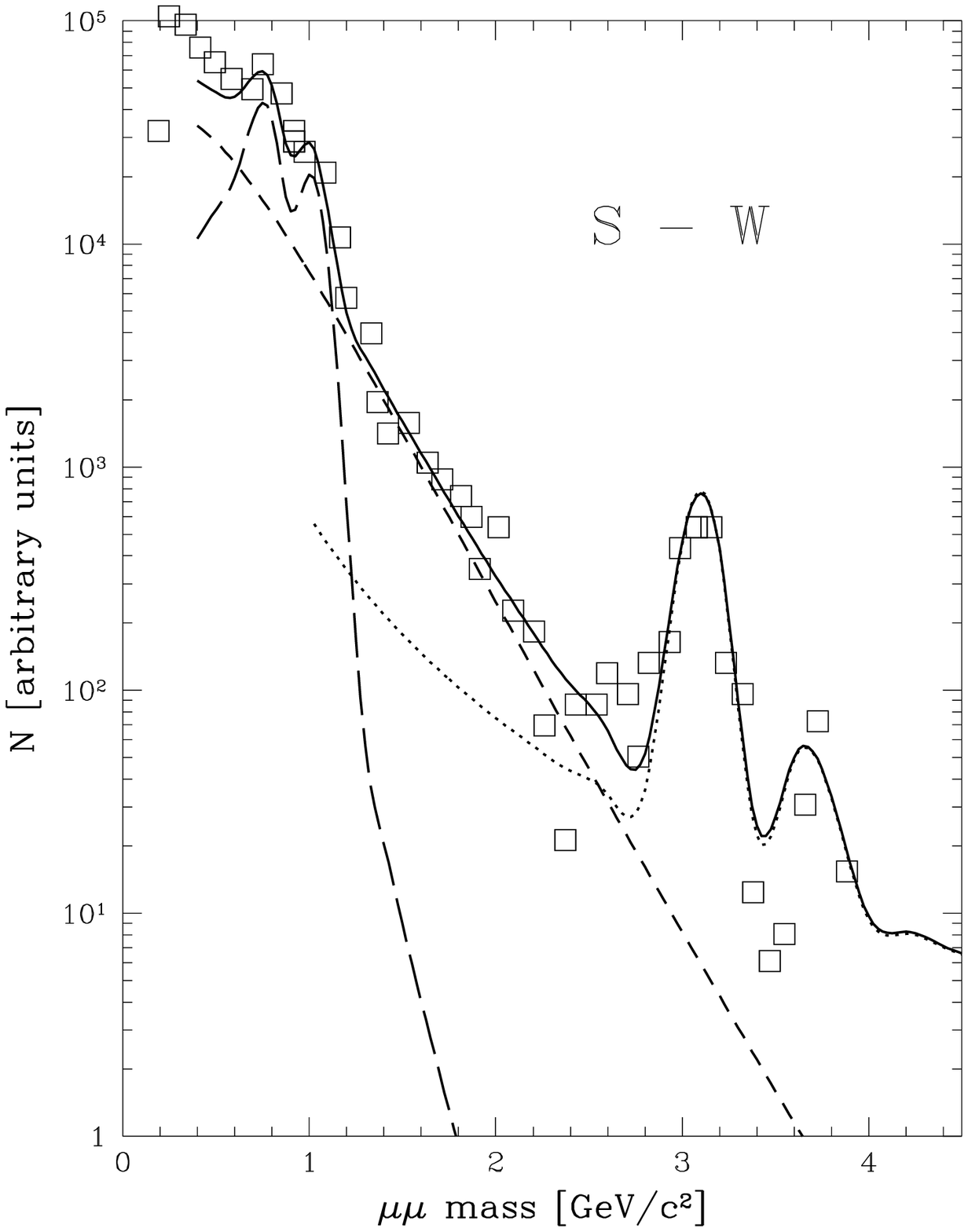}}
\vspace*{-5cm}
\centerline{\bf FIGURE 3}
\end{figure}
\end{document}